\documentclass[aps,prd,twocolumn,superscriptaddress, nofootinbib,floats,floatfix]{revtex4-1}
\usepackage{bm}
\usepackage{amsmath}
\usepackage{amsfonts}
\usepackage{aas_macros}
\usepackage{graphicx}
\usepackage[hidelinks]{hyperref}
\usepackage{booktabs}
\usepackage{color}
\usepackage[dvipsnames]{xcolor}
\usepackage[capitalise]{cleveref}
\usepackage[caption=false]{subfig}

\creflabelformat{equation}{#2\textup{#1}#3}

\begin{document}

\title{Constraints on quantum gravity and the photon mass from gamma ray bursts}

\author{D. J. Bartlett}
\email{deaglan.bartlett@physics.ox.ac.uk}
\affiliation{Astrophysics, University of Oxford, Denys Wilkinson Building, Keble Road, Oxford, OX1 3RH, United Kingdom}
\author{H. Desmond}
\affiliation{Astrophysics, University of Oxford, Denys Wilkinson Building, Keble Road, Oxford, OX1 3RH, United Kingdom}
\author{P. G. Ferreira}
\affiliation{Astrophysics, University of Oxford, Denys Wilkinson Building, Keble Road, Oxford, OX1 3RH, United Kingdom}
\author{J. Jasche}
\affiliation{The Oskar Klein Centre, Department of Physics, Stockholm University, AlbaNova University Centre, SE 106 91 Stockholm, Sweden}

\begin{abstract}
    Lorentz invariance violation in quantum gravity (QG) models or a nonzero photon mass, $m_\gamma$, would lead to an energy-dependent propagation speed for photons, such that photons of different energies from a distant source would arrive at different times, even if they were emitted simultaneously. By developing source-by-source, Monte Carlo-based forward models for such time delays from gamma ray bursts, and marginalising over empirical noise models describing other contributions to the time delay, we derive constraints on $m_\gamma$ and the QG length scale, $\ell_{\rm QG}$, using spectral lag data from the BATSE satellite. We find $m_\gamma < 4.0 \times 10^{-5} \, h \, {\rm eV}/c^2$ and $\ell_{\rm QG} < 5.3 \times 10^{-18} \, h \, {\rm \, GeV^{-1}}$ at 95\% confidence, and demonstrate that these constraints are robust to the choice of noise model. The QG constraint is among the tightest from studies which consider multiple gamma ray bursts and the constraint on $m_\gamma$, although weaker than from using radio data, provides an independent constraint which is less sensitive to the effects of dispersion by electrons.
\end{abstract}

\maketitle

\section{Introduction}

High energy astrophysical transients at cosmological distances allow us to test the fundamental assumptions of the standard models of cosmology and particles physics, such as the weak equivalence principle (WEP), Lorentz invariance (LI) or the massless nature of the photon (for a review, see \cite{Wei_2021}). If any of these assumptions are incorrect, photons of different energies propagate differently through spacetime, which could be observable in the spectral lags of gamma ray bursts (GRBs).

The hitherto elusive unification of quantum mechanics and general relativity is expected to exhibit so-called quantum gravity (QG) effects at energies of order the Planck scale, $E_{\rm P} \approx 1.2 \times 10^{19} {\rm \, GeV}$. By extension of the uncertainty principle, one may expect spacetime no longer to appear smooth on distance scales $\Delta x \sim 1 / E_{\rm P}$ \cite{Wheeler_1998}, and thus have a nontrivial refractive index for particles propagating through it. Hence, at low energies in QG theories, the photon velocity, $v$, is expected to depend on the energy, $E$, as \citep{AmelinoCamelia_1998}
\begin{equation}
    \label{eq:v quantum gravity}
    v \approx 1 - \xi \frac{E}{E_{\rm QG}},
\end{equation}
where $\xi = \pm 1$ and $E_{\rm QG}$ is the QG energy scale, constituting LI violation. Such a linear modification to $v$ is expected in a range of QG models \citep[see Section 1 of][and references therein]{Ellis_2019}. We dub the $\xi=+1$ and $\xi = -1$ models ``subluminal QG'' and ``superluminal QG'' respectively, based on the value of $v$ for nonzero $E$. One may expect that $\xi=+1$ otherwise photons would quickly lose energy due to gravitational \v{C}erenkov radiation \citep{Moore_2001}, however we consider both signs in this work since current constraints \cite{Kostelecky_2015} from \v{C}erenkov radiation only consider models which would lead to even powers of $E$ in \autoref{eq:v quantum gravity}. Superluminal photons would also decay into electron--positron pairs above a threshold energy, providing an independent test of such theories \citep{Klinkhamer_2008}.

The photon's dispersion relation could also be modified if it has a nonzero mass, $m_\gamma$, such that
\begin{equation}
    \label{eq:v massive photon}
    v = \sqrt{1 - \frac{m_\gamma^2}{E^2}}
        \approx 1 - \frac{1}{2} \frac{m_\gamma^2}{E^2},
\end{equation}
as can arise in theories with Lorentz and supersymmetry breaking \citep{Bonetti_2017b,Bonetti_2018}.

In this work we constrain these theories by considering the energy-dependent time delay between photon arrival time (spectral lag) of GRBs. For the majority of GRBs the spectral lag is positive, i.e., high energy photons are detected before lower energy photons. Although there are numerous mechanisms to explain this \citep[see e.g.][]{Shen_2005,Lu_2006,Daigne_2003,Peng_2011,Du_2019,Lu_2018,Uhm_2016}, this behaviour is qualitatively the same as for a massive photon or a QG theory with $\xi=+1$ and could thus provide evidence for such theories.

Constraints on $E_{\rm QG}$ from GRBs have previously been obtained using a variety of methods \citep{Ellis_2003,Ellis_2006,Ellis_2008,Rodriguez_2006,Bolmont_2008,Lamon_2008,Vasileiou_2013,MAGIC_2020,Wei_2017a,Wei_2017b,Wei_2017c,Tajima_2009,FLAT_2009,Chang_2012,Nemiroff_2012,Zhang_2015,Ellis_2019,Shao_2010,Xu_2016a,Xu_2016b,Xu_2018,Liu_2018,Agrawal_2021}, with the most stringent lower bounds of $E_{\rm QG} > 9.3 \times 10^{19} {\rm \, GeV}$ for the subluminal ($\xi=+1$) and $E_{\rm QG} > 1.3 \times 10^{20} {\rm \, GeV}$ for the superluminal ($\xi=-1$) models arising from GRB 090510 \citep{Vasileiou_2013}. In many cases these constraints are obtained using a handful of GRBs, do not propagate uncertainties in the redshifts of sources, or suffer from uncertain systematics in the model for other contributions to the spectral lag.

It is clear from \autoref{eq:v massive photon} that tighter constraints on $m_\gamma$ can be obtained using lower frequency photons,
and thus fast radio bursts (FRBs), pulsars and magnetars provide useful probes of $m_\gamma$ \cite{Bonetti_2016,Wu_2016,Wei_2021,Shao_2017,Wei_2018,Wei_2020,Xing_2019,Zhang_2016,Schaefer_1999,Bonetti_2017,Bentum_2017}.
However, the scaling with frequency ($v \propto E^{-2}$) is identical to the time delay due to dispersion by electrons \cite{Alonso_2021}, which is negligible for gamma rays, but leads to degeneracies at radio frequencies. The majority of constraints from radio frequencies neglect this important contribution \citep{Wei_2021}, although recently  \citep{Shao_2017,Wei_2018,Wei_2020} the plasma effect has been incorporated into a Bayesian analysis of FRBs, leading to constraints of $m_\gamma < 4 \times 10^{-15} {\rm \, eV}/c^2$. Slightly tighter constraints of $2.9\times 10^{-15} {\rm \, eV}/c^2$ have also been obtained \citep{Xing_2019}, but these do not include the plasma effect in the analysis. GRBs have previously been used to constrain the photon mass, but these either compare to radio frequencies \citep{Zhang_2016} or do not consider alternative causes for the time delay \citep{Schaefer_1999}.

There are two aims of this work. First, we address the issues highlighted when constraining $E_{\rm QG}$ by constructing probabilistic source-by-source forward models of the time delays of 448 GRBs from the BATSE satellite \citep{Hakkila_2007,Yu_2018} and marginalising over empirical models describing astrophysical and observational contributions to the measured time delay. These techniques were developed in \citet{Bartlett_2021_WEP} who, using the same sources as here, constrain WEP violation to $\Delta \gamma < 2.1 \times 10^{-15}$ at $1 \sigma$ confidence between photon energies of $25 {\rm \, keV}$ and $325 {\rm \, keV}$. Second, we provide constraints on the photon mass that are independent of radio observations and are thus insensitive to potential systematics in modelling the propagation of photons at radio frequency.

In \Cref{sec:Methods} we describe how we forward model the time delay, introduce our models for competing astrophysical and observational effects, and derive the likelihood function. We compare our predictions to the observations via a  Markov Chain Monte Carlo (MCMC) algorithm and present the results in \Cref{sec:Results}. \Cref{sec:Conclusions} concludes.

\section{Methods}
\label{sec:Methods}

\subsection{Forward modelling the time delay}

As noted by \citep{Jacob_2008}, for energy-dependent photon speeds, one cannot simply multiply the difference in photon velocity in the observer's frame by the distance travelled by a photon, but one must consider the cosmological redshift of the photon. This results in a time delay between photons of observed energy $E_i$ and $E_j$, $\Delta t_{ij} \equiv t_j - t_i$,
of
\begin{equation}
    \label{eq:td quantum gravity}
    \Delta t_{ij}^{\left( {\rm QG} \right)} = \xi \frac{E_i - E_j}{E_{\rm QG}} \int_{0}^z \frac{1 + z^\prime}{H \left( z^\prime \right)} {\rm d} z^\prime \equiv \xi \frac{\Delta E_{ij}}{E_{\rm QG}} I_{\rm QG},
\end{equation}
for the QG scenario, and
\begin{equation}
    \label{eq:td massive photon}
    \begin{split}
        \Delta t_{ij}^{\left( {\rm MP} \right)} &= \frac{m_{\gamma}^2}{2} \left(\frac{1}{E_i^2} - \frac{1}{E_j^2} \right)  \int_{0}^z \frac{1}{H \left( z^\prime \right) \left( 1 + z^\prime \right)^2} {\rm d}  z^\prime \\
        &\equiv \frac{m_{\gamma}^2}{2} \left(\frac{1}{E_j^2} - \frac{1}{E_i^2} \right) I_{\rm MP}
    \end{split}
\end{equation}
for the massive photon. We will assume a $\Lambda$CDM cosmology to determine the Hubble parameter $H(z)$; in general one should consider a variety of cosmological models \citep{Biesiada_2009,Pan_2015,Zou_2018,Pan_2020}, although this is beyond the scope of this work. In both cases we see that the parameter of interest ($E_{\rm QG}$ or $m_\gamma$) and the observed energy bands appear as scaling factors. We therefore simply need to compute the predicted theoretical time delay, $\Delta t_{ij}^{\left( {\rm th} \right)} \in \{ \Delta t_{ij}^{\left( {\rm QG} \right)}, \Delta t_{ij}^{\left( {\rm MP} \right)} \}$, using \Cref{eq:td quantum gravity,eq:td massive photon} for some fiducial $E_{\rm QG}$ or $m_\gamma$, and then rescale these parameters linearly for $E_{\rm QG}$ and quadratically for $m_\gamma$ (\Cref{eq:td quantum gravity,eq:td massive photon}) according to the $E_{\rm QG}$ or $m_\gamma$ being sampled.

We use the catalogue of 668 GRBs compiled by \citep{Yu_2018} from the BATSE satellite \citep{Hakkila_2007} since these not only have spectral lag data, but also pseudoredshifts calculated using the spectral peak energy-peak luminosity relation \cite{Yonetoku_2004}. Since the pseudo-redshift calibration only involves GRBs with redshifts below 4.5, we remove the 220 sources with pseudo-redshifts above this. We consider the adjacent pairs of the four energy channels of BATSE
(Ch1: 25-60 keV, 
Ch2: 60-110 keV, 
Ch3: 110-325 keV and 
Ch4: $>$325 keV),
i.e. $\left( i,j \right) \in \{ \left( 2,1 \right), \left( 3,2 \right), \left( 4,3 \right) \}$, where we neglect energy pairs where no lag is recorded.

Since the redshift values are uncertain, as in \citep{Bartlett_2021_WEP}, we draw $N_{\rm MC}=10^4$ redshifts per source from a two-sided Gaussian with upper and lower uncertainties equal to the uncertainties calculated in \citep{Yonetoku_2004}. Doubling $N_{\rm MC}$ yields identical results, indicating that the number of samples is adequate. For each sample we evaluate the integrals $I_{\rm QG}$ and $I_{\rm MP}$ using the \texttt{scipy.integrate} subpackage \cite{2020SciPy-NMeth}. To determine the appropriate $E_i$ to use in \Cref{eq:td quantum gravity,eq:td massive photon}, for each sample we draw an energy randomly from a distribution proportional to the best-fit spectral model for that GRB as given in the BATSE 5B Gamma-Ray Burst Spectral Catalog \citep{Goldstein_2013}. Furthermore, at each iteration we draw the parameters for the model from Gaussian distributions with means and widths as given in the catalogue.

The resulting $N_{\rm MC}$ samples are assumed to follow a Gaussian mixture model (GMM) \cite{scikit-learn}, such that the likelihood function for $\Delta t_{ij}^{\left( {\rm th} \right)}$ for some source $s$ is
\begin{equation}
    \label{eq:Integral likelihood}
	\mathcal{L}_{s} \left( \Delta t_{ij}^{\left( {\rm th} \right)} \right) = \sum_\alpha \frac{w^{\left( \alpha \right)}_{sij}}{\sqrt{2 \pi}\tau^{\left( \alpha \right)}_{sij}}
	\exp \left[ - \frac{\left( \Delta t_{ij}^{\left( {\rm th} \right)} - \lambda^{\left( \alpha \right)}_{sij}\right)^2}{2 {\tau^{\left( \alpha \right)}_{sij}}^2} \right],
\end{equation}
where
\begin{equation}
	\sum_\alpha w_{sij}^{\left( \alpha \right)} = 1, \quad  w_{sij}^{\left( \alpha \right)} \geq 0,
\end{equation}
and the sum runs over the number of Gaussian components. We compute an independent GMM for each source, and choose the number of Gaussians which minimise the Bayesian information criterion (BIC)
\begin{equation}
\label{eq:BIC}
	{\rm BIC} = \mathcal{K} \log \mathcal{N} - 2 \log \hat{\mathcal{L}},
\end{equation}
for $\mathcal{K}$ model parameters, $\mathcal{N}=N_{\rm MC}$ data points, and maximum likelihood estimate $\hat{\mathcal{L}}$.

\subsection{Modelling the noise}

Quantum gravity or a photon mass are not the only types of physics that can lead to spectral lags: these may also be generated through intrinsic differences in the emission of photons of different wavelength at the source or their propagation through the medium surrounding the GRB, or through instrumental effects at the observer. Without a robust physical model for the time delays these lead to, we model them using a generic functional form (a sum of Gaussians) with free parameters that we marginalise over in constraining $m_\gamma$ and $\ell_{\rm QG}$. We refer to any contribution to the time delays other than quantum gravity or a photon mass as ``noise.''

As demonstrated in \citep{Bartlett_2021_WEP}, for the WEP violation case one cannot accurately describe the noise by a single Gaussian. Instead, we model these extra contributions, $B_{ij}$, to the observed delay
\begin{equation}
    \Delta t_{ij}^{(\rm obs)} = \Delta t_{ij}^{(\rm th)} + B_{ij},
\end{equation}
as a sum of $N_{\rm G}$ Gaussians such that the likelihood for a given $B_{ij}$ is
\begin{equation}
    \label{eq:Noise likelihood}
    \mathcal{L} \left( B_{ij} \right) 
    = \sum_\beta \frac{\omega^{\left( \beta \right)}_{ij}}{\sqrt{2 \pi}\sigma^{\left( \beta \right)}_{ij}} \exp \left[ - \frac{\left( B_{ij} - \mu^{\left( \beta \right)}_{ij}\right)^2}{2 {\sigma^{\left( \beta \right)}_{ij}}^2} \right].
\end{equation}
where
\begin{equation}
	\sum_\beta \omega_{ij}^{\left( \beta \right)} = 1, \quad  \omega_{ij}^{\left( \beta \right)} \geq 0,
\end{equation}
and $\beta \in \{0, 1, \ldots, N_{\rm G} - 1 \}$. The components are labelled in order of decreasing weights, i.e. $\omega_{ij}^{\left( \beta \right)} \geq \omega_{ij}^{\left( \beta + 1 \right)}$.

In \autoref{eq:Noise likelihood} we have assumed that the noise only depends on the observed photon energies. Inspired by \citet{Ellis_2006}, we also consider noise models in which the means and widths of one or more of the Gaussians are redshift dependent,
\begin{equation}
    \mu^{\left( \beta \right)}_{ij} \to \mu^{\left( \beta \right)}_{ij} \left( 1 + z_{s} \right), \quad 
    \sigma^{\left( \beta \right)}_{ij} \to \sigma^{\left( \beta \right)}_{ij} \left( 1 + z_{s} \right),
\end{equation}
where $z_{s}$ is the quoted pseudoredshift of source $s$. These models capture an intrinsic contribution to the time delay from the source such as a ``magnetic-jet'' model for GRB emission \cite{Bosnjak_2012,Chang_2012}, whereas one would expect the redshift-independent models to describe observational effects.
By including these, we now have a wider range of noise models to choose from, increasing our confidence that the optimum model lies within this set.

\subsection{Likelihood model}

The likelihood of a given observed time delay, $\Delta t_{ij}^{(\rm obs)}$, for source $s$ is given by the convolution of \autoref{eq:Integral likelihood} (once we have appropriately scaled $\Delta t_{ij}^{\left( {\rm th} \right)}$ and the GMM parameters) and \autoref{eq:Noise likelihood},
\begin{equation}
    \begin{split}
    \mathcal{L}_{s} \left( \Delta t_{ij}^{(\rm obs)} \right) &=
    \sum_{\alpha \beta}
    \frac{w^{\left( \alpha \right)}_{s} \omega_{ij}^{\left( \beta \right)}}{\sqrt{2 \pi \left( {\tau}^{\left( \alpha \right)}_{sij}{}^2  + \sigma_{ij}^{\left( \beta \right)}{}^2 + \varepsilon_{sij}^2\right) }} \times \\
    & \exp \left[ - \frac{\left( \Delta t_{ij}^{(\rm obs)} - \lambda^{\left( \alpha \right)}_{sij} - \mu_{ij}^{\left( \beta \right)}  \right)^2}{2 \left( {\tau^{\left( \alpha \right)}_{sij}}{}^2 + \sigma_{ij}^{\left( \beta \right)}{}^2 + \varepsilon_{sij}^2 \right)} \right].
    \end{split}
\end{equation}
where we have also included the quoted measurement uncertainty in the spectral lag, $\varepsilon_{sij}$.

Assuming that all frequency pairs and sources are independent, the total likelihood for our dataset $\mathcal{D}$ is
\begin{equation}
    \mathcal{L} \left(\mathcal{D}| \bm{\theta}\right) = 
    \prod_{sij} \mathcal{L}_{s} \left( \Delta t_{ij}^{\left( {\rm obs} \right)} \right),
\end{equation}
where $\bm{\theta} \equiv \{ A, \mu_{ij}^{\left( \beta \right)}, \sigma_{ij}^{\left( \beta \right)}, \omega_{ij}^{\left( \beta \right)} \}$, and $A=m_\gamma$ or $\ell_{\rm QG} \equiv \xi E_{\rm QG}^{-1}$ depending on the theory considered. We choose to fit for the QG length scale, $\ell_{\rm QG}$, instead of $E_{\rm QG}$ since the infinite upper limit of the prior on $E_{\rm QG}$ becomes a zero lower limit on the prior for $\ell_{\rm QG}$. A separate set of noise parameters is fitted to each pair of frequencies, but we consider the target of interest, $A$, to be universal.

As in \citep{Bartlett_2021_WEP}, the deliberately wide priors, $P(\bm{\theta})$, in \autoref{tab:infered_parameter_summary} lead to difficulties in interpreting the Bayes ratio. To find the maximum likelihood and thus the BIC, we first optimise using the Nelder-Mead algorithm \citep{Gao_2012} with a simplex consisting of parameters drawn randomly from the prior. We repeat this ten times then compute the Hessian at the maximum likelihood point (MLP). Drawing 256 walkers from a Gaussian centred on the MLP with this Hessian, we run the \textsc{emcee} sampler \cite{emcee} for 10,000 steps to find a new estimate of the MLP using the $2.56 \times 10^6$ samples. If our estimate of the Hessian is not positive definite, we draw the walkers from log-normal distributions of unit width, centred on the MLP. We find that $\hat{\mathcal{L}}$ changes by less than 2 per cent for any $N_{\rm G}$ and for both theories considered if we only use the first 5,000 steps, which is much smaller than the change in BIC between different $N_{\rm G}$.

For computational convenience, we now use these posterior samples to restrict the size of the prior: we find the samples for which the change in $\chi^2$ ($\Delta \chi^2 = -2 \Delta \log \mathcal{L}$) from the MLP is 25 times the number of observed frequency pairs ($5\sigma$ for a Gaussian likelihood) and set the new prior such that it (just) encompasses these points. For some parameters we keep the prior wider than this to ensure that our results are not dominated by the choice of the prior. We now use Bayes' theorem to obtain the posterior distribution of $\theta$,
\begin{equation}
	\mathcal{P} \left(\bm{\theta}| \mathcal{D} \right) = \frac{\mathcal{L} \left(\mathcal{D}|\bm{\theta} \right) P \left( \bm{\theta} \right) }{\mathcal{Z}\left(\mathcal{D}\right)},
\end{equation}	
and evidence $\mathcal{Z}(\mathcal{D})$ with the nested sampling Monte Carlo algorithm MLFriends \citep{Buchner_2014,Buchner_2017} using the \textsc{UltraNest}\footnote{\url{https://johannesbuchner.github.io/UltraNest/}} package \citep{Buchner_2021}.
Since the prior is still treated as uniform and we do not use the Bayes ratio, reducing the size of the prior does not affect our results since it simply changes $\mathcal{P} \left(\bm{\theta}| \mathcal{D} \right)$ by a multiplicative constant except in regions where it is already negligible.

\begin{table}
    \caption{Priors on photon mass, QG length scale and parameters describing the empirical noise model (\autoref{eq:Noise likelihood}). All priors are uniform in the range given.
    }
    \label{tab:infered_parameter_summary}
    \centering
    \begin{tabular}{c|c}
    Parameter & Prior \\
    \hline
    $ m_\gamma \ / \ {\rm meV}/c^2$ & $\left[ 0, 1 \right]$\\ 
    $ \ell_{\rm QG} \ / \ {\rm GeV}^{-1}$ & $\left[ -10^{-14}, 10^{-14} \right]$\\ 
    $\mu_{ij}^{\left( \beta \right)} / \, {\rm s}$ & $\left[ -15, 15 \right]$\\
    $\sigma_{ij}^{\left( \beta \right)} / \, {\rm s}$ & $\left[ 0, 15 \right]$\\
    $\omega_{ij}^{\left( \beta \right)}$ & $[0, 1], \quad \sum_\beta \omega_{ij}^{\left( \beta \right)} = 1, \quad \omega_{ij}^{\left( \beta \right)} \geq \omega_{ij}^{\left( \beta + 1 \right)}$\\
    \end{tabular}
\end{table}

\section{Results}
\label{sec:Results}

\begin{figure*}
     \centering
    \subfloat[]{
         \centering
         \includegraphics[width=0.49\textwidth]{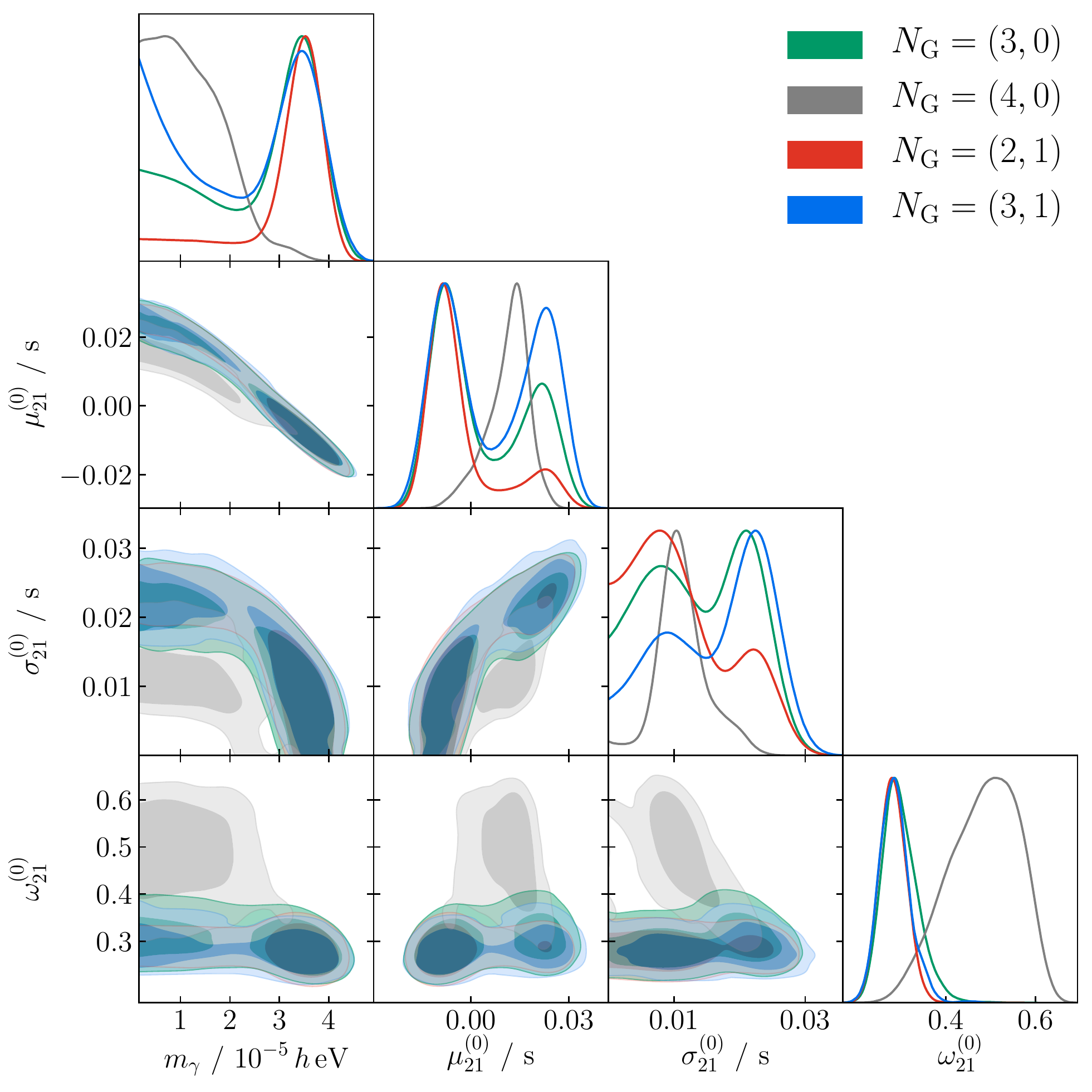}
         \label{subfig:Mass}
     }
     \subfloat[]{
         \centering
         \includegraphics[width=0.49\textwidth]{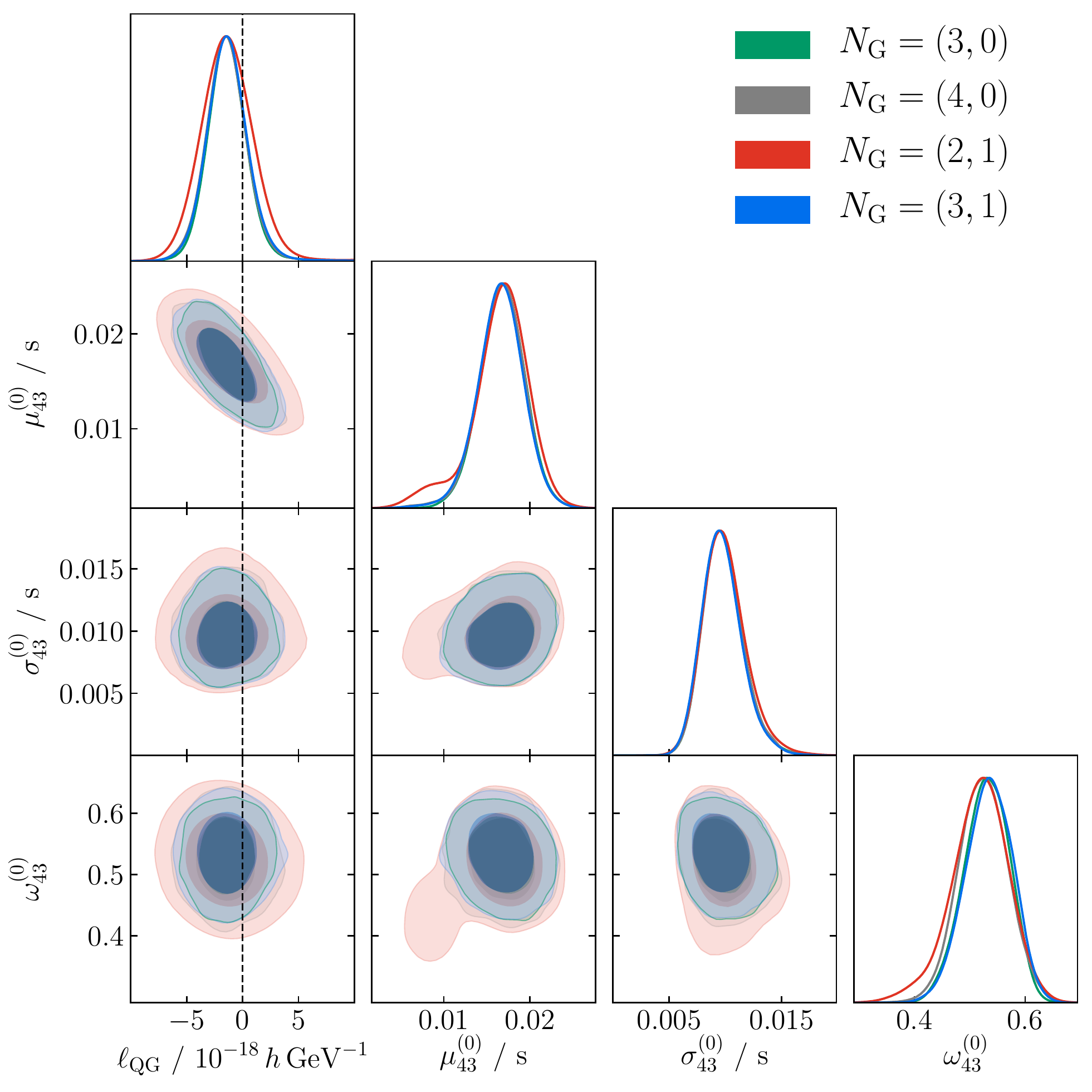}
         \label{subfig:Taylor}
     }
       \caption{Constraints on \protect\subref{subfig:Mass} the photon mass, $m_\gamma$, and \protect\subref{subfig:Taylor} the QG length scale, $\ell_{\rm QG}$, and the noise parameters of the Gaussian which is most correlated with the signal. The legend gives the number of redshift independent and dependent Gaussians used to describe the noise, respectively. A QG model with superluminal photon speed at high energies ($\xi = -1$) is defined to have a negative $\ell_{\rm QG}$. For the QG theories there is little degeneracy with the noise, whereas $m_\gamma$ is highly correlated with the mean of the highest-weighted Gaussian, making the constraint on $m_\gamma$ more sensitive to the noise model employed.
         }
		\label{fig:Posterior}
\end{figure*}

For both theories considered, we find that $N_{\rm G} = 3$ and 4 Gaussians have comparable BIC values, and that the best noise models contain either one redshift-dependent Gaussian or are completely independent of redshift, suggesting that observational effects dominate the noise. Using the \texttt{GetDist} package \cite{GetDist_2019}, in \autoref{fig:Posterior} we show the corner plots for $m_\gamma$ and $\ell_{\rm QG}$ and the parameters for the component of the noise model which is most correlated with the signal. We note that, from the energy dependence of \Cref{eq:v massive photon,eq:v quantum gravity}, it is unsurprising that $m_\gamma$ is most sensitive to the noise parameters from the frequency pair with the lowest energies, but for $\ell_{\rm QG}$ this is the pair with the largest range of energies.

We find that, for the QG theories, the results are relatively independent of the noise model, and the best-fit model gives a constraint of $\left| \ell_{\rm QG} \right| < 5.3 \times 10^{-18} \, h {\rm \, GeV^{-1}}$ at 95\% confidence, where $h \equiv H_0 / (100 {\rm \: km \: s^{-1} \: Mpc^{-1}})$ for Hubble constant $H_0$. We quote our results in terms of $h$ to remain agnostic as to the true value of $H_0$, given the ``Hubble tension'' between the value from the SH0ES collaboration ($H_0=73.1 \pm 1.3 {\rm \, km s^{-1} Mpc^{-1}}$ \cite{Riess_2020}) and that from Planck ($H_0=67.4\pm0.5$ \cite{Planck_2020}). The maximum of the marginalised one-dimensional posterior is, perhaps coincidentally, in the same direction as \cite{Ellis_2008} (accounting for the different sign in the definition), indicating a slight preference for a superluminal QG theory. For the photon mass inference we see that $m_\gamma$ is correlated with the highest weighted Gaussian for frequency pair $(i,j) = (2,1)$ and that the marginalised one-dimensional posterior is more sensitive to the noise model. In all cases we find $m_\gamma < 4.0 \times 10^{-5} \, h {\rm \, eV}$ at 95\% confidence. 

Besides the clear proportionality with $h$, we find that our constraints are relatively independent of cosmological parameters; varying $\Omega_{\rm m0}$ in the range $0.25-0.35$ with fixed $h$ changes the constraints by $\lesssim 10\%$.

We previously assumed that the uncertainty on the pseudoredshift of a source can be described by a two-tailed Gaussian. To test the impact of this assumption, we run the analysis assuming zero redshift error, and find the constraints on $m_\gamma$ and $\ell_{\rm QG}$ tighten by 18\% and 3\% respectively. 
Due to the pseudoredshift calibration, we removed all GRBs with pseudoredshifts above $z_{\rm max} =4.5$. Increasing this to $z_{\rm max} = 6$ slightly tightens the constraints by 3\% for the massive photon case and 5\% for the QG theories. If we include all GRBs from \citep{Yu_2018} then the constraint on $m_\gamma$ is again virtually unchanged, whereas we find a nonzero (at $3\sigma$ confidence) value of $\ell_{\rm QG} = -1.6 \times 10^{-18} \, h {\rm \, GeV^{-1}}$. We find that this ``detection'' is driven by two GRBs (\mbox{4B 910619} and \mbox{4B 921112-)} at $z\sim 7.5$ which have negative time delays. Upon excising these potential outliers, the constraint is again consistent with zero at $1\sigma$ confidence.

\section{Conclusions}
\label{sec:Conclusions}

In this work we have considered two theories in which the photon propagation speed depends on energy: a quadratic correction due to nonzero photon mass, $m_\gamma$, and a quantum gravity scenario in which the photon speed depends linearly on energy, as is expected in a wide range of models. By forward-modelling the expected time delays of photons of different frequencies for a large sample of GRBs, we find constraints on the photon mass of $m_\gamma < 4.0 \times 10^{-5} \, h \, {\rm eV}$ and on the QG length scale of $|\ell_{\rm QG} |< 5.3 \times 10^{-18} \, h \, {\rm \, GeV^{-1}}$ at 95\% confidence. Our constraints on $m_\gamma$ are significantly less stringent than those from radio observations, however are much less sensitive to the effects of dispersion by electrons, which has the same frequency dependence in the dispersion relation as a massive photon.

A large number of previous attempts to constrain QG with the spectral lag of GRBs assume a simple noise model in which the non-QG contribution to the time delay is proportional to $(1+z)$ and is constant for all sources, even though \citet{Ellis_2006} demonstrated that ignoring stochasticity dramatically changes the results. Moreover, these studies often only use a small sample of GRBs (sometime only one), but one requires a statistical sample to provide trustworthy constraints. Our constraints are among the tightest astrophysical constraints on QG which use multiple sources (see Table 1 of \cite{Wei_2021}) and we have demonstrated that these are robust to how one models other astrophysical and observational contributions to the spectral lag. Our constraints are comparable to \citet{Ellis_2019}, who use the irregularity, kurtosis and skewness of GRBs instead of spectral lag to find $E_{\rm QG} \equiv \left| \ell_{\rm QG} \right|^{-1} \gtrsim {\rm few} \times 10^{17} {\rm \, GeV}$.

It is expected that detecting GRBs at $>$100 ${\rm \, GeV}$ should be routine in the future \cite{Zhang_2019}; with more, higher energy measurements one should begin to probe $E_{\rm QG}$ near the Planck energy, $E_{\rm P}$. Since one expects $E_{\rm QG} \sim E_{\rm P}$, either a nonzero or null detection of LI violation at these scales will significantly constrain which QG theories are allowed. With very few other known tests of quantum gravity, it is therefore important that future work should develop more theoretically motivated noise models for GRB spectral lag than we have used here to ensure that any detection or rejection of new physics is not due to incorrect modelling of the astrophysical processes governing GRB emission.

\acknowledgements
{
D.J.B. is supported by STFC and Oriel College, Oxford. H.D. is supported by St John's College, Oxford. P.G.F. is supported by the STFC. H.D. and P.G.F. acknowledge financial support from ERC Grant No 693024 and the Beecroft Trust. J.J. acknowledges support by the Swedish Research Council (VR) under the project No. 2020-05143 -- ``Deciphering the Dynamics of Cosmic Structure".
This work was done within the Aquila Consortium (\url{https://www.aquila-consortium.org/}).
}

\bibliographystyle{apsrev4-1}
\bibliography{references}

\end{document}